\documentclass{article}
\usepackage{epsf}
\usepackage{amsfonts}
\usepackage{times}

\newcommand{\ket}[1]{\left | \, #1 \right \rangle}

\newcommand{\bra}[1]{\left \langle #1 \, \right |}
\newcommand{\proj}[1]{\ket{#1}\!\!\bra{#1}}

\newcommand{\Tr}{{\mathrm{Tr}}}

\begin{document}
\vspace{1in}

\begin{center}
{\Large The Classical Capacity Achievable by a Quantum Channel
Assisted by Limited Entanglement
}\\*[2ex]
{\large Peter W. Shor}\\
{Dept.\ of Mathematics} \\
Massachusetts Institute of Technology\\
Cambridge, MA 02139, USA\\*[3ex]
Dedicated to Alexander S. Holevo on the occasion of his 60th birthday.
\end{center}

\noindent
{\bf Abstract:} 
We give the trade-off curve showing the capacity of a quantum
channel as a function of the amount of entanglement used by the
sender and receiver for transmitting information.  The endpoints
of this curve are given by the Holevo-Schumacher-Westmoreland
capacity formula and the entanglement-assisted capacity, which
is the maximum over all input density matrices of the quantum
mutual information.  The proof we give is based on the 
Holevo-Schumacher-Westmoreland formula, and also
gives a new and simpler proof for the entanglement-assisted 
capacity formula. 

\section{Introduction}
Information theory says that the capacity of a classical channel is
essentially unique, and is representable as a single numerical quantity,
giving the amount of information that can be transmitted asymptotically 
per channel use \cite{Shannon48,Cover}. Quantum channels, unlike
classical channels, do not have a single numerical quantity which can
be defined as their capacity for transmitting information.  Rather, 
quantum channels appear to have at least four different natural 
definitions of capacity, depending on the auxiliary resources allowed, 
the class of protocols allowed, 
and whether the information to be transmitted is classical or quantum.

This paper will discuss the transmission of {\em classical} information
over quantum channels.  One of the first results in this area was an
upper bound proved by Holevo \cite{Holevobound}
on how much classical information could be transmitted by an ensemble
of quantum states.  
Holevo \cite{Holevo98} and Schumacher and Westmoreland \cite{SW97}
independently discovered proofs that
this bound was achievable.  This gives the theorem

{\bf Theorem} (Holevo; Schumacher and Westmoreland):
{\em The classical capacity obtainable using codewords composed of 
tensor products of signal states~$\sigma_{i}$, where the probability of 
using $\sigma_i$ is $p_i$, is
\begin{equation}
\chi(\{\sigma_i,p_i\}) = 
H(\sum_i p_i \sigma_i) - \sum_i p_i H(\sigma_i),
\label{formula-chi}
\end{equation}
where $H(\rho) = -\Tr \rho \log \rho$ is the von Neumann entropy of
the density matrix $\rho$.}

Note that $\chi(\{\sigma_i, p_i\})$ is a function of the probabilistic 
ensemble of signal states $\{\sigma_i, p_i\}$ that we have chosen, where 
state $\sigma_i$ has $p_i$.  When it is clear what this ensemble
is, we may simply denote this by $\chi$.  

A memoryless quantum communication channel is
a linear trace preserving completely positive map.  
Such maps can be expressed as
\[
{\cal N}(\rho) = \sum_i A_i \rho A_i^\dag,
\]
where the $A_i$ satisfy $\sum_i A_i^\dag A_i = I$.
A natural guess at the capacity of a quantum channel ${\cal N}$ would
be the maximum of $\chi$ over all possible 
probability distributions of channel outputs,
that is, the capacity would be
\begin{eqnarray}
\label{def_chi_max}
\chi_\mathrm{max} ({\cal N}) &=& \max_{\{\sigma_i, p_i\}} 
\chi \big(\{{\cal N}(\sigma_i), p_i\}\big)\\
&=& \max_{\{\sigma_i, p_i\}} 
H({\cal N}(\sum_i p_i \sigma_i)) - \sum_i p_i H({\cal N}(\sigma_i))
\nonumber
\end{eqnarray}
since the sender can effectively communicate to the receiver any
of the states ${\cal N}(\sigma_i)$.  
This maximum can be achieved using pure states $\sigma_i$. 
This quantity is clearly achievable.   
We do not know whether this is
the capacity of a quantum channel; this is reducible to the
question of additivity of the quantity $\chi_\mathrm{max}$, 
\begin{equation} 
\chi_\mathrm{max}({\cal N}_1 \otimes {\cal N}_2) \stackrel{?}{=} 
\chi_\mathrm{max}({\cal N}_1) + \chi_\mathrm{max}({\cal N}_2),
\end{equation}
a question
which has in recent years received much study \cite{AHW,MSW,Shor-add,
King1,Holevo-add-constrained}.  
If we require the protocols
to send states that are tensor products on the different
uses of the quantum channel, this is indeed the achievable
capacity.   However, 
if the use of entanglement between separate inputs to the channel helps
to increase channel capacity,  it would be possible to exceed
this $\chi_\mathrm{max}$.  The capacity of a quantum channel can be
shown to be the regularized form of Eq.~(\ref{def_chi_max}), that is,
\begin{equation}
\lim_{n \rightarrow \infty}
\frac{1}{n} \chi_\mathrm{max} ({\cal N}^{\otimes n}) .
\end{equation}

The next capacity we discuss is the entanglement-assisted 
capacity of a quantum channel \cite{CEshort,CElong}.
In the entanglement-assisted capacity, the sender and 
receiver share entanglement at the start of the protocol, which they are 
allowed to use in the communication protocol.  
The entanglement-assisted capacity is given by the following formula.

{\bf Theorem} (Bennett, Shor, Smolin, Thapliyal):
{\em The classical capacity obtainable using a quantum channel
${\cal N}$ is
\begin{equation}
C_E = \max_\rho \ \ H(\rho) + H ( {\cal N}(\rho) ) - 
H\Big(({\cal N}\otimes I)(\phi_\rho)\Big)
\label{formula-EAC}
\end{equation}
where $\phi_\rho$ is a state over the tensor product of the
input space and a reference system, ${\cal H}_\mathrm{in} \otimes {\cal
H}_\mathrm{ref}$, whose reduced density matrix on the channel's
input space is $\rho$, 
i.e., $\Tr_2 \phi_\rho = \rho$. }  

The amount
of pure state entanglement consumed by the protocol given in $\cite{CElong}$ 
can be shown to be asymptotically $H(\rho)$ ebits per channel use, where
$\rho$ is the density matrix maximizing Eq.~(\ref{formula-EAC}), and
an ebit is the amount of pure state entanglement in an EPR pair of qubits.  

This naturally leads to several questions. Is $H(\rho)$ ebits per channel
use the amount of entanglement required to achieve the
entanglement-assisted capacity?  More generally, if the amount of
entanglement available is $P < H(\rho)$ ebits
per channel use, how much classical information
can be transmitted?  We answer these questions in the following theorem.

{\bf Theorem 1.} 
{\em
If the available entanglement per channel use is restricted to $P$ ebits,
there is a protocol achieving the information rate given by
\begin{eqnarray} \nonumber
 \max_{\{\rho_i, p_i\}} &&
\sum_i p_i H (\rho_i)
+H \big( {\cal N} (\sum_i p_i \rho_i)\big)
-\sum_i p_i H \big( ({\cal N} \otimes {\cal I})
(\phi_{\rho_i})\big)\\
 \mathrm{subject\ to} &&\sum_i p_i H(\rho_i) \leq P,
\label{formula-tradeoff}
\end{eqnarray}
where $\Tr_2 \phi_{\rho_i} = \rho_i$.
Here, the maximization is over all probabilistic ensembles of density matrices
$\{\rho_i, p_i\}$ where $\rho_i \in {\cal H}_\mathrm{in}$, 
$\sum_i p_i = 1$, and the average entropy of the ensemble,
$\sum_i p_i H(\rho_i)$, is at most $P$.}  

In the  case where $P=0$, this gives the Holevo capacity $\chi_\mathrm{max}$
of Eq.~(\ref{def_chi_max}), as the $\rho_i$ must all be pure states. 
In the case where $P$ is sufficiently large, this gives the 
entanglement-assisted capacity $C_E$ of Eq.~(\ref{formula-EAC}).
Since we do not know whether the Holevo capacity is additive, 
we clearly cannot show that the above capacity trade-off 
is additive; this is an open question.  
We can however prove that this formula is an upper bound
if we restrict ourself to protocols where the sender
and receiver start by sharing pure entangled quantum states,   
and the sender is not allowed to distribute one of these entangled
states among more than one channel use, the same restriction under
which we know the Holevo capacity $\chi$ is the correct formula
for unassisted classical capacity.  To get the true
capacity trade-off formula (if it is not additive), we may
have to regularize this formula.  That is, to take the limit of the
normalized entanglement-assisted capacity for the channel 
${\cal N}^{\otimes n}$ as $n$ goes to infinity.

This theorem can also be derived using the methods of \cite{Brady}. 
However, we give a quite different and somewhat simpler proof than 
in \cite{Brady} for
the trade-off formula, as well as a simpler proof than \cite{CElong} for the 
entanglement-assisted capacity.
This proof relies on the Holevo-Schumacher-Westmoreland theorem above, so
in this paper we are showing that knowing the left endpoint of
this trade-off curve lets us derive the entire curve.

\section{The Protocol}

We now give the protocol that asymptotically achieves the capacity
(\ref{formula-tradeoff}).
We use block coding.
We will let $n$ be the number of channel uses in our 
block coding protocol.  This protocol will take $n$ entangled states and
use them as the input for these channel uses.  It will not distribute
one of these entangled states over more than one channel use, but it
will permute the entangled states before sending them through
the channels, so the mapping of the entangled states
to the channel inputs depends on the message being sent.
Suppose that the maximum of Eq.~(\ref{formula-tradeoff})
occurs at the ensemble
$\{\rho_i,p_i\}$.  
We assume that the sender and the receiver start by sharing a number 
$n$ of entangled states
where there are $n_i \approx n p_i$ states for which the reduced 
density matrix is $\rho_i$.  For the proof that our protocol
achieves its desired capacity, we will use the
Holevo-Schumacher-Westmoreland theorem with $n!\, 2^{(d-1)n}$ 
signal states, where $d$ is the
dimension of the input space to the channel.  
These signal states are described as follows.  

First, 
Alice applies  to her part of the state $\ket{\phi_{\rho_i}}$ a 
random
sign change $\pm 1$ to the phase of each of the eigenvalues of $\rho_i$.  
Note that there are $2^{d-1}$ possible phase changes for each of the
states $\ket{\phi_{\rho_i}}$, as
Alice can without loss of generality apply the phase $+1$ to the first 
eigenvalue (since an overall phase change
does not alter the quantum state).  
Next, 
Alice applies a random permutation
to the $n$ entangled states she shares with Bob.  
Since there are $n!$ permutations, 
we have $n!\, 2^{n(d-1)}$ signal states total.  

Before we can continue, we need a lemma.  

{\bf Lemma 1.} {\it
Suppose we have $n$ density matrices, $\rho_1$, $\rho_2$, $\ldots$,
$\rho_n$, which are drawn at random from some probability
distribution on
density matrices.  Then 
\begin{equation}
\lim_{n \rightarrow \infty}
\frac{1}{n} \mathrm{E} \left[ 
H\Big( \frac{1}{n!} \sum_\pi 
\rho_{\pi(1)} \otimes \rho_{\pi(2)} \otimes \cdots \otimes
\rho_{\pi(n)}\Big)\right]
= H(\bar{\rho})
\label{lemma1-eq}
\end{equation}
where the sum is over all $n!$ permutations $\pi$ of the $n$ density matrices, 
the expectation $\mathrm{E}$ is over the random choice of
$\rho_1$ \ldots $\rho_n$, and $\bar{\rho}$ is the average density
matrix for the probability distribution that the $\rho_i$ are drawn from.
}

That the left hand side of Eq.~(\ref{lemma1-eq}) is at most 
the right hand side
follows immediately from the subadditivity of entropy of quantum states.
The proof of the other direction will be deferred until later.

The proof of Theorem 1 is slightly nicer if we let the $n_i$ be random
variables obtained by drawing $n$ density matrices from a distribution where
$\rho_i$ occurs with probability $p_i$.  In other words, instead of 
Alice and Bob starting each coding block with exactly $n_i$ copies
of the entangled state $\ket{\phi_{\rho_i}}$, 
they use the next $n$ states in a sequence of
shared states where $\ket{\phi_{\rho_i}}$ occurs with probability $p_i$.  It
is not hard to prove that the protocol also works when they start with
exactly $n_i \approx n p_i$ states, although we will not prove this in
the paper.

We now look at the signal states more carefully.  The first term in 
the Holevo capacity $\chi$, Eq.~(\ref{formula-chi}), is the
entropy of the average output signal received by Bob.  This signal consists
of two parts, the quantum state $A$ which was originally held by
Alice, and was subsequently modified and sent through $n$ uses
of the channel ${\cal N}$, and the quantum state $B$, which
Bob originally held and has kept.  

The random phase change applied by Alice
disentangles Alice and Bob's
entangled states $\ket{\phi_i}$.  We will work in the basis of 
the eigenvalues of $\rho_i$.   Let these eigenvalues be $\ket{v_{ij}}$.
In this basis, 
$\ket{\phi_{\rho_i}} = \sum_j \sqrt{\lambda_{ij}} \ket{v_{ij}}\ket{v_{ij}}$. 
After the random phase change, the density matrix is 
$\sum_j \lambda_{ij} \proj{v_{ij}}\otimes \proj{v_{ij}}$.
This is the same density matrix as is given by the ensemble
containing the state
$\proj{v_{ij}}\otimes \proj{v_{ij}}$ with probability $\lambda_{ij}$.
Let us assume then that
Alice and Bob started by sharing
$n$ unentangled quantum states, each of which was in
the state $\ket{v_{ij}}\ket{v_{ij}}$ with probability $p_i\lambda_{ij}$.  
We will bound the entropy of Bob's average signal state by using 
this second ensemble, which must give the same answer, as the entropy
depends only on the density matrix.
What we do is add an extra, classical, variable, which we denote
by $T$.  We let $T$ tell us the type class of the distribution; that
is, the variable $T$ holds the numbers $n_i$ and 
the numbers $m_{ij}$,
where $\sum_j m_{ij} = n_i$ and where $m_{ij}$ tells how many of these
quantum systems started in the state $\ket{v_{ij}}\ket{v_{ij}}$.  
The reason we
do this is that
after Alice applies the phase changes and the random permutation
to her quantum states, if we condition on $T$
the quantum states Alice and Bob hold
are now independent.  This is because 
Alice inputs into the channel a mixture of all permutations of the $n$ states
consisting of $m_{ij}$ copies of 
$\ket{v_{ij}}$ for each $i,j$, and this mixed state is determined solely 
by $T$.  By entropy inequalities and the definition of conditional
entropy.  
\begin{eqnarray}
\nonumber
H(A) + H(B) \geq H(AB) & \geq & H(ABT) - H(T) \\
\nonumber
&=& H(AB | T) \\
&=& H(A | T) + H(B | T) \\
\nonumber
&=& H(AT) + H(BT) - 2H(T) \\
\nonumber
&\geq & H(A) + H(B) - 4H(T) .
\end{eqnarray}
However, since $H(T) = O(\log n)$, we need only
estimate $H(A)$ and $H(B)$ to compute the asymptotics of $H(AB)$.  
We have
\begin{equation} 
H(B) = \sum_i n_i H(\rho_i) \approx n \sum_i p_i H(\rho_i).
\end{equation}
The state $A$ is a mixture of all permutations of the density matrices
${\cal N}(\proj{v_{ij}})$, where ${\cal N}(\proj{v_{ij}})$ occurs 
with probability $p_i
\lambda_{ij}$, so by Lemma 1, 
\begin{equation} 
H(A) \approx nH({\cal{N}}(\bar{\rho}))
\end{equation} 
where
$\bar{\rho} = \sum_i p_i \rho_i$.  Thus, the
first term of the HSW formula, $H(AB)$, is approximately 
\begin{equation} 
H(AB) \approx H(A) + H(B) \approx
n\left(\sum_i p_i H(\rho_i) + H({\cal{N}}(\sum_i p_i \rho_i))\right).
\end{equation}

Finally, we look at the second term in the Holevo capacity $\chi$,
Eq.~(\ref{formula-chi}).  
This is the entropy $({\cal N}^{\otimes n}\otimes I)(\proj{\Phi_{\pi,P}})$, 
where 
$\ket{\Phi_{\pi,P}}$ is the signal state her half of which
Alice inputs into the channel.  
This state was produced by
Alice first performing a random phase change $P$ 
in the eigenbasis of $\rho_i$, 
to her half of all of ther quantum states $\ket{\phi_{\rho_i}}$,
and then applying a random
permutation $\pi$ to all $n$ of her states.  It is easy to check that if Bob
knows what these random phase changes and permutation were, 
he can undo them.  Thus, 
all $n!\,2^{(d-1)n}$ signal
states give rise to the same joint entropy, which is 
$\sum_i n_i H\big((I\otimes {\cal N}) (\proj{\phi_{\rho_i}})\big)$.  
This is the last term of Eq.~(\ref{formula-tradeoff}).
We thus have a protocol that asymptotically
achieves Eq.~(\ref{formula-tradeoff}).

\newpage
\section{Proof of the Lemma}

We now prove the following lemma, which will imply Lemma 1.

{\bf Lemma 2.}
{\it 
Suppose that we have $n$ density matrices
$\rho_1$, $\rho_2$, $\ldots$, $\rho_n$.  Let
\begin{equation}
\bar{H}(\bar{\rho}_k) =
\frac{1}{n!} \sum_\pi H\big( {\textstyle\frac{1}{k} (\rho_{\pi(1)} + \ldots + 
\rho_{\pi(k)})}\big)
\end{equation}
be the expected entropy of the average of $k$ of these density matrices 
chosen randomly without replacement 
from the $n$ density matrices.  Then}
\begin{equation}
H \left( \frac{1}{n!} \sum_\pi \rho_{\pi(1)} \otimes \rho_{\pi(2)} \otimes 
\ldots \otimes \rho_{\pi(n)} \right) \geq
\sum_{k=1}^n \bar{H}(\bar{\rho}_k)
\label{lemma-2-eq}
\end{equation}

{\bf Proof:}
We let $T_k$ be a variable which gives the values of the images of the first 
$k$ elements of the permutation $\pi$: $\pi(1)$, $\pi(2)$, $\ldots$, $\pi(k)$.
Then
\begin{eqnarray} 
\label{telescoping}
&& \hspace*{-.7in}
H\bigg(\frac{1}{n!} \sum_\pi \bigotimes_{j=1}^{k+1} \rho_{\pi(j)}\bigg) -
H\bigg(\frac{1}{n!} \sum_\pi \bigotimes_{j=1}^{k} \rho_{\pi(j)}\bigg) \\
\nonumber
&\geq &
H\bigg(\frac{1}{(n-k)!} \sum_{\pi|T_k} \bigotimes_{j=1}^{k+1}
\rho_{\pi(j)}\bigg) -
H\bigg(\frac{1}{(n-k)!} \sum_{\pi|T_k} \bigotimes_{j=1}^{k}
\rho_{\pi(j)}\bigg) \\
\nonumber
&=& H\bigg(\frac{1}{(n-k)!} \sum_{\pi|T_k} \rho_{\pi(k+1)}\bigg)\\
\nonumber
&=& \bar{H}(\bar{\rho}_{(n-k)}),
\end{eqnarray}
where $\pi|T_k$ is the set of permutations which have their first $k$ 
elements fixed
by $T_k$.   The inequality above is an application of the
strong superadditivity property of quantum entropy.  Now,
by adding the left hand sides of the above expression (\ref{telescoping})
for $k$ between $0$ and $n-1$,
we obtain a telescoping series which gives the left hand side of 
Eq.~(\ref{lemma-2-eq}).
Adding the right-hand side of Eq.~(\ref{telescoping}) for $k$
between $0$ and $n-1$ gives the right
hand side of Eq.~(\ref{lemma-2-eq}), proving the lemma.

Suppose now that $\rho_1$, $\rho_2$, $\ldots$, $\rho_n$ are matrices drawn
identically and independently
from some probability distribution.  The above lemma implies that
\begin{equation}
\mathrm{E} \left[ 
H\Big( \frac{1}{n!} \sum_\pi 
\rho_{\pi(1)} \otimes \rho_{\pi(2)} \otimes \cdots \otimes
\rho_{\pi(n)}\Big)\right]
\geq \sum_{k=1}^n \mathrm{E} \bar{H}(\bar{\rho}_k)
\end{equation}
where now note that 
$\mathrm{E} \bar{H} ( \bar{\rho}_k )$ is the 
expected entropy of the average of 
$k$ density matrices drawn from the probability distribution.
But for a finite dimensional quantum space, 
$\mathrm{E} \bar{H}(\bar{\rho}_k)$ is easily seen to converge to 
$H(\bar{\rho})$, where 
$\bar{\rho}$ is the average density matrix of the probability distribution.
This completes the proof of Lemma 1.
\section{The upper bound.}

What we do now is show an upper bound on the capacity of a quantum
channel assisted
by limited entanglement of $P$ ebits per channel use, subject to 
the proviso that Alice cannot
input a state entangled over more than one channel use.  We will assume the
following scenario.  Alice and Bob start with a set of pure entangled states
$\ket{\phi_i}$, where we define
$\rho_i = \mathrm{Tr}_B \proj{\phi_i}$.  We let Alice
perform an arbitrary unitary transformation on her states, and then send part 
(or all) of the resulting state through the channel.  We do not let Alice 
input the
same $\ket{\phi_i}$ into more than one channel use.  We will use the
Holevo bound to bound the information that
Bob can receive using such a protocol.

We will first start with the assumption that Alice's part of their shared 
state
occupies a Hilbert space with the same dimension as the channel input, 
i.e., $\dim \rho_i = \dim {\cal H}_\mathrm{in}$.  Now, Alice will perform a 
unitary transformation to obtain the state $U_j \rho_i U_j^\dag$, and send it
through the channel.  We now use the Holevo bound, Eq.~(\ref{formula-chi}),
to bound the capacity Alice and Bob can achieve using such a protocol.  
Again, Bob's signal consists of the state he received from Alice through
the channel together with the state that he kept.  By the
subadditivity of entropy, the
first term of Eq.~(\ref{formula-chi}) is bounded by the entropy of the 
average output
of Alice's channel plus the average entropy of the reduced states held by Bob.
If $U_j \rho_i U_j^\dag$ is sent with probability $p_{ij}$, then this first 
term is
bounded by
\[
\sum_{i,j} p_{ij} H(\rho_i) + 
H\big(\sum_{i,j} p_{ij} {\cal N}(U_j \rho_i U_j^\dag)\big),
\]
which is the same as the first two terms in the formula
(\ref{formula-tradeoff}), assuming 
that
we used the ensemble $\{U_j\rho_i U_j^\dag, p_{ij}\}$ in formula
(\ref{formula-tradeoff}).  The
second term of the Holevo
bound (\ref{formula-chi}) is also identical in these two scenarios.  
Specifically, it is
\begin{equation}
\sum_{ij} p_{ji} H\Big(({\cal N}\otimes I)\big((U_j\otimes I)\proj{\phi_{i}}
(U_j^\dag \otimes I)\big)\Big) =
\sum_{ij} p_{ji} H\Big(({\cal N}\otimes I)(\tau_{ij})\Big),
\end{equation}
where $\tau_{ij} = \proj{\phi_{U_j\rho_iU_j^\dag}}$ is a purification 
of $U_j \rho_i U_j^\dag$.  Thus, if
Alice applies unitary transformations to $\rho_{ij}$ and inputs the entire 
resulting
state into the channel, she cannot achieve a capacity better than that given
by Theorem 1.

We now show that the same bound applies if Alice is allowed to put only a 
part of her quantum state through the channel.  That is, Alice and Bob
share an entangled pure state $\ket{\phi_i}$ where 
$\Tr_B \proj{\phi_i} = \rho_i
\in {\cal H}_\mathrm{in} \otimes {\cal H}_\mathrm{ref}$.
Alice puts ${\cal H}_\mathrm{in}$ through the channel, and discards
${\cal H}_\mathrm{ref}$.
We will compare the capacity achieved by this
case with that achieved by an alternative scenario.  If Alice measures the
reference system ${\cal H}_\mathrm{ref}$  in the basis determined by the 
eigenvalues of 
$\mathrm{Tr}_\mathrm{in} \rho_i$, she obtains a probability distribution 
$p_{ij}$ over 
quantum states $\rho_{ij} \in {\cal H}_\mathrm{in}$.  We will show that 
replacing
$\rho_i$ with the ensemble of states $\{\rho_{ij}, p_{ij}\}$ increases the 
capacity, while 
decreasing the amount of pure state entanglement consumed by the protocol.

Holevo's bound shows that the capacity of such a protocol can be at most
\begin{equation}
\sum_i p_i H(\rho_{i}) + H\Big({\cal N}(\sum_i \Tr_\mathrm{ref} \rho_i)\Big) 
+ \sum_i p_i 
H\Big(({\cal N} \otimes I)(\Tr_\mathrm{ref} \proj{\phi_{\rho_i}})\Big),
\label{formula-encodings}
\end{equation}
where $\phi_{\rho_i}$ is the joint pure state of Alice and Bob.  If all
the $\rho_i$ are in ${\cal H}_\mathrm{in}$ (so we need no reference system
${\cal H}_\mathrm{ref}$), then (\ref{formula-encodings}) gives the same
capacity as (\ref{formula-tradeoff}).  

Let the output of the channel be denoted by $A$, and let
the reference system after
Alice's measurement (which now contains the classical variable $j$ telling 
which of the measurement outcomes was obtained) be $R$.  Finally, let the 
part of the entangled state $\ket{\phi_{i}}$ held by Bob be $B$.  
The replacement of $\rho_i$ by
the ensemble $\{\rho_{ij}, p_{ij}\}$ does not change the average output of the
channel, so the second term of (\ref{formula-encodings})
giving the entropy of the average output of the channel is
unchanged.  The contribution to the first and third terms 
from the state $\rho_i$ is proportional to
\begin{equation}
H(B) - H(AB).
\label{ssa1}
\end{equation}
Replacing $\rho_i$ by the ensemble $\{\rho_{ij}, p_{ij}\}$ gives a 
contribution proportional to
\begin{equation}
\big( H(BR) - H(R) \big) - \big( H(ABR) - H(R) \big).
\label{ssa2}
\end{equation}
This second contribution (\ref{ssa2}) is larger than the first 
(\ref{ssa1}) by the property of strong 
subadditivity
of quantum entropy.

It is also easy to see that the amount of pure state entanglement consumed by
the protocol decreases after the replacement of $\rho_i$ by
$\{\rho_{ij}, p_{ij}\}$, since Alice and
Bob can obtain the ensemble of states $\{\rho_{ij}, p_{ij}\}$
from the state $\rho_i$ using solely local quantum operations
and classical communication (LOCC operations), and these never increase the
amount of entanglement.  We thus see that the assumption that all of Alice's
part of the entangled states was sent through the channel did not impose any
restrictions on channel capacity.

Finally, let us note that if Alice takes her parts of two different 
entangled pure 
states and sends them through
one channel use, this also cannot increase the capacity 
of the protocol.  
To see this, note that this case has essentially already 
been taken into account in our analysis, as the tensor product of 
the two pure entangled states can be considered as a single 
entangled state.  Thus, the only case this is not covered by our analysis
is when Alice's channel inputs are entangled over more than one channel 
use.  This case is discussed briefly in the next section.

\section{Discussion}
We have given a formula that tells how much the classical capacity
of a quantum channel can be increased by the use of a limited amount
of entanglement between the sender and receiver, which is
consumed by the protocol for transmitting information. 
This paper 
is quite different in approach than the paper \cite{Brady}, 
which also gives a proof for this trade-off curve.  
It also yields
a simpler proof of the original entanglement-assisted
capacity formula \cite{CElong}.

It is not known whether we need to regularize the trade-off
formula to find the capacity.  In light of the recent 
discovery that many of the additivity problems in quantum information 
theory are equivalent \cite{Shor-add}, a natural question is to
ask whether this is equivalent to these other problems.  We have not
been able to show this, although it is clearly at least as hard,
since additivity of the Holevo capacity, which is one of the equivalent
problems, is the special case of the trade-off curve when no
entanglement is consumed.  

Finally, let us note
that in order to achieve the capacity formula (\ref{formula-tradeoff})
without Alice using inputs entangled between different channel uses, 
it appears that Alice and Bob need to be able to start by sharing arbitrary
pure entangled states $\ket{\phi_{\rho_i}}$, and that 
it is not sufficient for Alice and Bob to start by sharing solely EPR pairs.
However, since pure state entanglement is an interconvertible resource
\cite{LP}, if we remove the restriction on Alice sending states entangled
between different channel uses, then we can use EPR pairs for the shared
entanglement consumed by the protocol.

\small
\setlength{\itemsep}{0pt}
\setlength{\parskip}{0pt}
\setlength{\parsep}{0pt}


\begin{thebibliography}{99}
\setlength{\itemsep}{0pt}
\setlength{\parsep}{0pt}
\setlength{\parskip}{0pt}
\bibitem{AHW} G. G. Amosov, A. S. Holevo, R. F. Werner, ``On some additivity
problems in quantum information theory,'' {\em Problems in information
transmission,\/} vol. 36, pp. 25--34, 2000;  
arXiv e-print math-ph/0003002.

\bibitem{CEshort} C. H. Bennett, P. W. Shor, J. A. Smolin and A. V. Thapliyal,
``Entanglement-assisted classical capacity of noisy quantum channels,''
{\em Phys. Rev. Lett.,\/} vol. 83, pp. 3081--3084, 1999.

\bibitem{CElong} C. H. Bennett, P. W. Shor, J. A. Smolin and A. V. Thapliyal,
"Entanglement-assisted capacity of a quantum channel and the reverse 
Shannon theorem",  {\em IEEE Trans. Info. Theory,} vol. 48, pp. 2637--2655,
2002;  
arXiv e-print quant-ph/0106052.

\bibitem{Cover} T. M. Cover and J. A. Thomas, {\em Elements of Information
Theory,} Wiley, New York, 1991.

\bibitem{Brady} I. Devetak, A. W. Harrow, and A. Winter, ``A family
of quantum protocols,'' arXiv e-print quant-ph/0308044.

\bibitem{Holevobound} A. S. Holevo, ``Information theoretical aspects of
quantum measurements,'' {\em Probl. Info. Transm. (USSR),\/}
vol. 9, no. 2, pp. 31--42, 1973 (in Russian); 
[translation: A. S. Kholevo, {\em Probl. Info. Transm.,\/}
vol. 9, pp. 177--183, 1973].

\bibitem{Holevo98} A. S. Holevo, ``The capacity of the quantum channel with
general signal states,'' {\em IEEE Trans. Info. Theory,\/} vol. 44, 
pp. 269--273, 1998.

\bibitem{HolEA1} A. S. Holevo, ``On entanglement-assisted classical
capacity,'' {\em J. Math.\ Phys.\/} vol.~43, pp. 4326--4333 (2002);
arXiv e-print quant-ph/0106075.

\bibitem{HolEA2} A. S. Holevo, ``Entanglement-assisted capacity of constrained
channels,'' 
arXiv e-print quant-ph/0211170.

\bibitem{Holevo-add-constrained}, A. S. Holevo and M. E. Shirokov,
``On Shor's channel extension and
constrained channels,'' arXiv e-print quant-ph/0306196.

\bibitem{King1} C. King, ``The capacity of the quantum depolarizing channel,''
{\em IEEE Trans.\ Inform.\ Theory,\/} vol.. 49, pp. 221--229, 2003;
arXiv e-print quant-ph/0204172.

\bibitem{LP} H.-K. Lo and S. Popescu, ``The classical communication cost
of entanglement manipulation: Is entanglement an inter-convertible
resource?'' {Phys.\ Rev.\ Lett.} {\bf 83}, pp. 1459--1462, 1999.

\bibitem{MSW} K. Matsumoto, T. Shimono and A. Winter, ``Remarks on additivity
of the Holevo channel capacity and of the entanglement of formation,
arXiv e-print quant-ph/0206148.

\bibitem{SW97} B. Schumacher and Westmoreland,
``Sending classical information via a noisy quantum channel,''
{\em Phys. Rev. A,\/} vol. 56, pp. 131--138, 1997.

\bibitem{Shannon48} C. E. Shannon, ``A mathematical theory of communication,''
{\em The Bell System Tech. J.,\/} vol. 27, pp. 379--423, 623--656, 1948.

\bibitem{Shor-add} P. W. Shor, ``Additivity of the classical capacity
of entanglement-breaking channels,'' {\em J. Math. Physics,\/} vol. 43,
pp. 4334-4340 (2002).


\end{thebibliography}
\end{document}